\documentclass[twocolumn]{revtex4}
\usepackage{epsfig}
\usepackage{amsmath}
\usepackage{amsfonts}
\usepackage{amssymb}

\begin{document}

\title{A Floquet analysis perspective of driven light-matter interaction models}
\author{Jonas Larson}
\email[e-mail address: ]{Jonas.Larson@fysik.su.se}
\affiliation{Department of Physics, Stockholm University, SE-106 91 Stockholm, Sweden}

\date{\today}

\begin{abstract}
In this paper, we analyze the harmonically driven Jaynes-Cummings and Lipkin-Meshkov-Glick models using both numerical integration of time-dependent Hamiltonians and Floquet theory. For a separation of time-scales between the drive and intrinsic Rabi oscillations in the former model, the driving results in an effective periodic reversal of time. The corresponding Floquet Hamiltonian is a Wannier-Stark model, which can be analytically solved. Despite the chaotic nature of the driven Lipkin-Meshkov-Glick model, moderate system sizes can display qualitatively different behaviors under varying system parameters. Ergodicity arises in systems that are neither adiabatic nor diabatic, owing to repeated multi-level Landau-Zener transitions. Chaotic behavior, observed in slow driving, manifests as random jumps in the magnetization, suggesting potential utility as a random number generator. Furthermore, we discuss both models in terms of what we call Floquet Fock state lattices.
\end{abstract}
\maketitle

\section{Introduction}
The Jaynes-Cummings (JC) model stands as a cornerstone in our explorations of light-matter interaction within the realm of quantum optics ~\cite{jaynes1963comparison,shore1993jaynes}. Despite its apparent simplicity, the model's profound impact was evidenced nearly four decades ago when experimental validation showcased its efficacy, see for example the review~\cite{larson2022jaynes}. This model, depicting a two-level system (representing matter, often an atom) coupled to a harmonic oscillator (typically a single mode of the electromagnetic field), has since served as a catalyst for deeper explorations into quantum dynamics. One might expect that only under extreme conditions would the JC model accurately encapsulate the complexities of tailored light-matter interaction. However, its validation in experimental settings and its generalizations to more intricate models have solidified its significance.

Central to the JC model and similar models in quantum optics is the limitations few degrees of freedom alongside certain symmetries. The adoption of the {\it rotating wave approximation} (RWA) further enhances the model's utility, ensuring the preservation of the number of excitations within the system. This makes such models ideal for probing fundamental questions within quantum mechanics~\cite{haroche2006exploring}. Furthermore, the introduction of classical driving forces can result in novel phenomena.  Again, either a rotating frame is considered, or an RWA is applied, leading to an effective time-independent Hamiltonian. For certain drives, however, an RWA might not be justified, and one must consider the full time-dependent problem. 

Early investigations into periodically driven JC models, such as that by Schlicher~\cite{schlicher1989jaynes}, gave insight into more complex scenarios. In this vein, we extend our analysis to encompass sinusoidal coupling within the JC model and explore its implications. Additionally, we delve into the sinusoidally driven {\it Lipkin-Meshkov-Glick} (LMG) model~\cite{lipkin1965validity,dusuel2004finite,ribeiro2007thermodynamical,morrison2008dynamical,larson2010circuit,latorre2005entanglement}, broadening our scope to encompass diverse manifestations of light-matter interaction.

To facilitate our study, we employ {\it Floquet theory}~\cite{holthaus2015floquet,eckardt2017colloquium,rudner2020floquet}, a tool for tackling time-dependent quantum problems. Floquet theory enables the transformation of such problems into equivalent time-independent formulations, albeit within an extended state space termed the Floquet space. While numerically challenging due to the infinite nature of this space, Floquet analysis offers invaluable insights into the system's dynamics.

In this work, we introduce the {\it Floquet Fock state lattice}, an extension of the {\it Fock state lattice} (FSL) to the Floquet space. The Floquet FSL serves as a representation of the Floquet Hamiltonian, shedding light on system evolution and revealing phenomena such as Bloch oscillations and level repulsion, indicative of chaos.

The paper is organized as follows: In the next section, we provide an overview of Floquet theory, particularly as applied to harmonic driving, where the Floquet Hamiltonian can be easily determined. Insec.~\ref{jcsec}, we examine the driven JC model, focusing on relatively slow drives where the driving acts as an effective time reversal. The subsequent section~\ref{lmgsec} addresses the LMG model, where we conduct an analysis of the time-dependent problem, discussing various manifestations of chaos. Additionally, we describe the time evolution in the Floquet FSL. Finally, we conclude in sec.~\ref{con} with a summary. Appendix~\ref{appA} presents the basics of FSLs and how the Floquet FSL emerges.

\section{Floquet theory for a harmonical drive}
The analogue of the Bloch theorem for time-periodic Hamiltonians is the Floquet theorem~\cite{shirley1965solution}. However, in quantum mechanics, time and space play distinct roles, as time is not an ordinary observable that can be assigned an operator. Nonetheless, as we will illustrate, it is possible to establish a correspondence of plane waves for the time 'coordinate' and incorporate the time-dependence of the Hamiltonian into a new (Floquet) Hamiltonian, residing in an expanded space comprising both the original Hilbert space and an additional Floquet space denoted as $\mathcal{H}\otimes\mathcal{T}$. This approach reveals that time evolution can be separated into slow and fast-scale components. The Floquet Hamiltonian serves as the time-evolution operator for the slow dynamics occurring over a period $T$, while the short time-scale evolution within one period is encapsulated by a {\it stroboscopic kick operator}. Although this topic is not commonly addressed in standard textbooks, several comprehensive introductions to the field of Floquet theory are available, as exemplified by references~\cite{holthaus2015floquet,eckardt2017colloquium,rudner2020floquet}. We present fundamental concepts here and direct interested readers to the aforementioned references for a more in-depth exploration.

The starting point is a periodic Hamiltonian $\hat H(t+T)=\hat H(t)$, with $T$ the period. The theorem states that there exists an orthonormal basis $\{|\psi_n(t)\rangle\}$, which under time-evolution of the Hamiltonian $\hat H(t)$, obeying 
\begin{equation}\label{floqstate0}
    |\psi_n(t+T)\rangle=e^{-i\varepsilon_n T}|\psi_n(t)\rangle.
\end{equation}
The states $|\psi_n(t)\rangle$ are called the {\it Floquet states}, $\varepsilon_n$ are the {\it quasi-energies}, and we use $\hbar=1$ throughout. Given the Floquet states, the strength of the theorem is that a general initial state $|\psi(0)\rangle$ will follow the simple evolution
\begin{equation}
    |\psi(t)\rangle=\sum_na_n|\psi_n(t)\rangle,
\end{equation}
where, importantly, the coefficients $a_n$ are time independent. To find the Floquet states we note that they can be decomposed into a time-periodic state multiplied be a phase factor as\footnote{We could shift the phase factor, $\exp[-i\varepsilon_nt]\rightarrow\exp[-i\varepsilon_n(t-\tau_0)]$ for some $\tau_0$, and it would modify the definition of the Floquet Hamiltonian. This is a gauge freedom we have.}
\begin{equation}\label{floqstate}
    |\psi_n(t)\rangle=e^{-i\varepsilon_n t}|\phi_n(t)\rangle,\hspace{1.2cm}|\phi_n(t+T)\rangle=|\phi_n(t)\rangle.
\end{equation}
By substituting~(\ref{floqstate}) into the Schr\"odinger equation one derives
\begin{equation}\label{floqham1}
\begin{array}{c}
    \left[\varepsilon_n+i\frac{d}{dt}\right]|\phi_n(t)\rangle=\hat H(t)|\phi_n(t)\rangle\\
    \Leftrightarrow\\
    \hat H^\mathrm{F}|\phi_n(t)\rangle\equiv\left[\hat H(t)-i\frac{d}{dt}\right]|\phi_n(t)\rangle=\varepsilon_n|\phi_n(t)\rangle.
\end{array}
\end{equation}
Here we have introduced the (time-independent) {\it Floquet Hamiltonian} $\hat H_\mathrm{F}$. Furthermore, owing to the periodicity of the Hamiltonian, we can Fourier expand the periodic states as
\begin{equation}\label{fexp}
    |\phi_n(t)\rangle=\sum_{m=-\infty}^\infty|\varphi_n(t),m\rangle\equiv \sum_{m=-\infty}^\infty e^{im\omega t}|\varphi_n,m\rangle,
\end{equation}
where $\omega=2\pi/T$, and $|\varphi_n,m\rangle$ is the $m$'th Fourier coefficient of $|\phi_n(t)\rangle$. We note that $|\varphi_n(t),m\rangle=e^{im\omega t}|\varphi_n(t),0\rangle$, and moreover 
\begin{equation}\label{timedis}
    \cos(\omega t)|\varphi_n(t),m\rangle=\frac{1}{2}\left[|\varphi_n(t),m+1\rangle+|\varphi_n(t),m-1\rangle\right].
\end{equation}
Thus, $\exp(\pm i\omega t)$ acts as an effective raising/lowering operator for the states $|\varphi_n,m\rangle$. It is clear that the Floquet state~(\ref{floqstate}) is invariant under the shifts $\varepsilon_n\rightarrow\varepsilon_n+m'\omega$ and $|\varphi_n(t),m\rangle\rightarrow|\varphi_n(t),m-m'\rangle$. As a result, the quasi-energies $\varepsilon_n$ can be restricted to any interval of length $\omega$, say $-\omega/2\leq\varepsilon_n<\omega/2$, which serves as the analogue of a Brillouin zone in frequency rather than in momentum space. 

To solve the full problem we still need to find $\varepsilon_n$ and $|\phi_n(t)\rangle$. Making use of the Fourier expansion~(\ref{fexp}) in eq.~(\ref{floqham1}) gives
\begin{equation}\label{floqeq2}
    \left(\sum_{m'=-\infty}^\infty H^{(m-m')}+m\omega\right)|\varphi_n,m'\rangle=\varepsilon_n|\varphi_n,m\rangle,
\end{equation}
with $H^{(m)}$ the Fourier coefficient of the Hamiltonian, i.e. $\hat H(t)=\sum_{m=-\infty}^\infty e^{im\omega t}H^{(m)}$. Equation~(\ref{floqeq2}) can be put on a matrix form $\hat H_\mathrm{F}|\varphi_n\rangle=\varepsilon_n|\varphi_n\rangle$, with the Floquet Hamiltonian expressed in the $m$-basis as
\begin{equation}
    \hat H^\mathrm{F}=\left[
    \begin{array}{cccc}
    \ddots & H^{(-1)} & H^{(-2)} & \\
    H^{(1)} & H^{(0)}-m\omega & H^{(-1)} & H^{(-2)}\\
    H^{(2)} & H^{(1)} & H^{(0)}-(m+1)\omega & H^{(-1)}\\
     & H^{(2)} & H^{(1)} & \ddots
    \end{array}\right].
\end{equation}
On the diagonal, $H^{(0)}$ is the time-averaged Hamiltonian. Given that the dimension of the Hilbert space is $d$, each time-independent block-Hamiltonian $H^{(m)}$ is also of dimension $d$. The dimension of the Floquet Hamiltonian is infinite since $m$ runs over all integers, and for numerics an $m$-truncation must be considered. What we have achieved is to transform the time-dependent problem into a time-independent eigenvalue problem, which solved provides the quasi-energies $\varepsilon_n$ and the Fourier coefficients for $|\phi_n(t)\rangle$. With these, the Floque states are obtained from eq.~(\ref{floqstate}).

The aforementioned separation of time-scales is not transparent from the discussion so far. First we note that for one period, the Floquet state~(\ref{floqstate}) is multiplied by a phase factor determined by the quasi-energy. Thus, in this case we can express the time-evolution operator over one period, and starting at $t_0$, in terms of the Floquet Hamiltonian as
\begin{equation}
    \hat U(t_0+T,t_0)=e^{-i\hat H^\mathrm{F}T}.
\end{equation}
In principle, the Floquet Hamiltonian can depend on the (gauge) choice $t_0$. It can be shown\footnote{This actually defines the operator $\hat K^\mathrm{F}$, given the Floquet Hamiltonian $\hat H^\mathrm{F}$ and the gauge choice.}~\cite{holthaus2015floquet}, that the full time-evolution operator could be split into two parts
\begin{equation}\label{timeop}
    \hat U(t,t_0)=e^{-i\hat K^\mathrm{F}t}e^{-i\hat H^\mathrm{F}t},
\end{equation}
where $\hat K^\mathrm{F}$ is sometimes referred to as {\it stroboscopic kick operator}. It obeys $e^{-i\hat K^\mathrm{F}(t+T)}=e^{-i\hat K^\mathrm{F}t}$, and given this partitioning of the time-evolution operator we identify $\hat H^\mathrm{F}$ as generating the slow time evolution over a full period, while $\hat K^\mathrm{F}$ provides the short time evolution occurring within one period. The above result is the Floquet theorem expressed in terms of the time-evolution operator. 

Hence, we have re-written the time-dependent problem, characterized by a time-evolution operator $\hat U(t,0)=T\exp\left[-i\int_0^t\hat H(t')dt'\right]$, with $T$ the time-ordering operator, into an effectively time-independent problem~(\ref{timeop}). The new problem is to be solved in the extended {\it Floquet space} that is comprised by the original Hilbert space $\mathcal{H}$ and the space $\mathcal{T}$ of the time-periodic states $|\phi_n(t)\rangle$, i.e. $\mathcal{F}=\mathcal{H}\otimes\mathcal{T}$. While time-independent problems are typically far more easy to solve than time-dependent ones, we pay the price that we are now dealing with a larger state space. Numerically, even if the dimension of $\mathcal{H}$ is finite, we face an infinite state space since the Fourier expansion~(\ref{fexp}) runs from  $-\infty$ to $+\infty$, and a truncation must be introduced~\cite{rudner2020floquet}. Actually, it is not uncommon to find that numerically solving the original time-dependent problem is more time-efficient than solving the time-independent Floquet problem. Nevertheless, this might give some valuable insight into the dynamics. 

In the examples to follow we consider harmonic driving characterized by some frequency $\omega$, i.e. $\hat H(t)=\hat H\left(\cos(\omega t)\right)$. There is a recipe how to construct the Floquet Hamiltonian in this case~\cite{klimov2009group}. To this end we note that the cosine function couples the $m$-state to the $(m\pm1)$-states in eq.~(\ref{timedis}). Thus, in the extended space spanned by the $\{|\varphi,m\rangle\}$ basis, $\cos(\omega t)$ is a tri-diagonal matrix with $1/2$ on its first off-diagonals. For brevity, if we write $|n,m\rangle\equiv|\varphi_n,m\rangle$ and introduce the ladder operators $\hat E^\pm|n,m\rangle=|n,m\pm1\rangle$, eq.~(\ref{timedis}) can be compactly written as 
\begin{equation}
    \left(\hat E^++\hat E^-\right)|n,m\rangle=|n,m+1\rangle+|n,m-1\rangle.
\end{equation}
The ladder operators, complemented with the diagonal operator $\hat E_0|n,m\rangle=m|n,m\rangle$, form what is called a {\it Euclidean algebra} that obey the commutation relations $\left[\hat E_0,\hat E^\pm\right]=\pm\hat E^\pm$ and $\left[\hat E^-,\hat E^+\right]=0$~\cite{klimov2009group}. Furthermore, we may note
\begin{equation}
    \hat U_{0}(t)\hat E^\pm\hat U_{0}^\dagger(t)=\hat E^\pm e^{\mp i\omega t},
\end{equation}
where $\hat U_{0}(t)=\exp\left(-i\hat E_0\omega t\right)$. Let us introduce the phase states $\hat E^-|\theta\rangle=e^{i\theta}|\theta\rangle$, and the time-dependent state $|\theta(t)\rangle=\hat U_0(t)|\theta\rangle$. We can then construct a Floquet Hamiltonian $\hat H^\mathrm{F}$ (in the extended space $\mathcal{F}$) such that the time-dependent Hamiltonian (in the Hilbert space $\mathcal{H}$) is $\hat H(t)=\langle\theta(t)|\hat H^\mathrm{F}|\theta(t)\rangle$. Note that the phase states $|\theta(t)\rangle\propto\sum_me^{-i\theta m}|m\rangle$ are also eigenstates of $\hat E^++\hat E^-$ with eigenvalues $2\cos(\theta)$, and we have a freedom to choose the phase, for example $\theta=0$.

Give the above, the recipe how to construct this Floquet Hamiltonian follows from the substitution
\begin{equation}\label{fmap}
    \cos(\omega t)\rightarrow\frac{\hat E^++\hat E^-}{2},\hspace{1.4cm}-i\frac{d}{dt}\rightarrow\omega\hat E_0,
\end{equation}
and hence
\begin{equation}
    \hat H^\mathrm{F}=\hat H\left(\left(\hat E^++\hat E^-\right)/2\right)+\omega\hat E_0.
\end{equation}

Before proceeding, let us define number, or Fock, states $|n\rangle$ in the general way, namely as the eigenstates of the number operator. Apart from the standard ones bosonic and fermionic ones, we complement the these with spin and Floquet states;
\begin{equation}\label{fockst}
    \begin{array}{lll}
     \mathrm{Bosonic}: &    &\hat n=\hat a^\dagger\hat a,\,\,\,\,\,\hat n|n\rangle=n|n\rangle,\,\,\,\,n\in\mathbb{Z}^+,  \\
     \mathrm{Fermionic}: &    &\hat n=\hat c^\dagger\hat c,\,\,\,\,\,\hat n|n\rangle=n|n\rangle,\,\,\,\,n=0,\,1,  \\ 
      \mathrm{Spin}: &    &\hat S_z=\frac{\hat S^+\hat S^--\hat S^-\hat S^+}{2},\,\,\,\,\,\hat S_z|m_s\rangle=m_s|m_s\rangle,\\
      \mathrm{Qubit}: &    &\hat\sigma_z=2\hat\sigma^+\hat\sigma^--\mathbb{I},\,\,\,\,\,\hat\sigma_z|s\rangle=s|s\rangle,\,\,\,\,s=\pm1,  \\
      \mathrm{Floquet}: &    &\hat E_0,\,\,\,\,\,\hat E_0|m\rangle=m|m\rangle,\,\,\,\,m\in\mathbb{Z}.  \\
    \end{array}
\end{equation}
Note that given the properties of the Euclidean algebra, its number operator $\hat E_0$ cannot simply be expressed in terms of its ladder operators $\hat E^\pm$.

\section{Driven Jaynes-Cummings model}\label{jcsec}
The Jaynes-Cummings model, comprising a two-level system, referred to as an atom, coherently coupled to a single boson mode, playing the role of the quantized electromagnetic field, is given by~\cite{jaynes1963comparison,shore1993jaynes,larson2022jaynes}
\begin{equation}
    \hat H_\mathrm{JC}=\frac{\Delta}{2}\hat\sigma_z+g(t)\left(\hat a^\dagger\hat\sigma^-+\hat a\hat\sigma^+\right).
\end{equation}
Here, $\Delta$ is the atom-field detuning, $g(t)$ is the light-matter coupling, $\hat a^\dagger/\hat a$ are the photon creation/annihilation operators obeying $\left[\hat a,\hat a^\dagger\right]=1$, and the $\hat\sigma$-operators are the Pauli $z$- and raising/lowering matrices, i.e. $\left[\hat\sigma^+,\hat\sigma^-\right]=\hat\sigma_z$ and $\left[\hat\sigma^\pm,\hat\sigma_z\right]=\mp2\hat\sigma^\pm$. We denote the eigenstates of the $z$ Pauli matrix as $|g\rangle$ (ground) and $|e\rangle$ (excited).

\begin{figure}[ht!]
\centering\includegraphics[width=8.3cm]{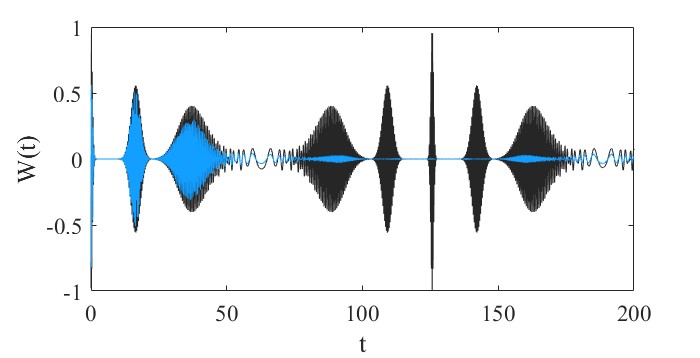}
\caption{The atomic inversion  as given in eq.~(\ref{atinv}) shown as the black curve, and the corresponding numerical results calculated from the truncated Floquet Hamiltonian~(\ref{wsham}) for a lattice of 1401 sites (light blue). The other parameters are $\omega=0.025$ and $g=1$, and the initial photon mode is in a coherent state with $\alpha=5$, and the atom in $|g\rangle$. At $t_\mathrm{tr}=\pi/2\omega$ the evolution is effectively reversed to result in a perfect revival at $t=\pi/\omega$. The discrepancy between the two results derives from the Floquet truncation; the larger lattices, the more accurate results at later times. }\label{fig1}
\end{figure}

The Hamiltonian can be derived from a minimal-coupling Hamiltonian under a series of approximations, see~\cite{larson2022jaynes}. Most important for us is the RWA that accounts for neglecting the counter-rotating terms $\hat a^\dagger\hat\sigma^+$ and $\hat a\hat\sigma^-$. In doing so, the number of excitations $\hat N=\hat n+\frac{1}{2}\hat\sigma_z$, with $\hat n=\hat a^\dagger\hat a$, is preserved such that the model possess a continuous $U(1)$ symmetry and the Hamiltonian is $2\times2$ block-diagonal in the Fock basis $\left\{|n,g\rangle,|n,e\rangle\right\}$\footnote{Each block $\hat h_n$ constitutes a $2\times2$ matrix apart from the vacuum $|0,g\rangle$ that is decoupled from any other states in the RWA.}. Thus, for each excitation $N$ sector we find the block Hamiltonian, spanned by the states $|n,e\rangle$ and $|n+1,g\rangle$, as
\begin{equation}\label{tjc}
    \hat h_n(t)=\left[
    \begin{array}{cc}
    \Delta/2 & g_0\cos(\omega t)\sqrt{n}\\
    g_0\cos(\omega t)\sqrt{n} & -\Delta/2
    \end{array}
    \right],
\end{equation}
where we assumed $g(t)=g_0\cos(\omega t)$. Let us for simplicity consider the case $\Delta=0$ where it is straightforward to decouple the two equations via a Hadamard transformation with the action $\hat\sigma_z\leftrightarrow\hat\sigma_x$. With $|n\rangle_{x\pm}=\left(|n-1,e\rangle\pm|n,g\rangle\right)/\sqrt{2}$ (and $|0\rangle_x=|0,g\rangle$) the eigenstates of the above Hamiltonian we can write down the time-evolved state for a general initial state as
\begin{equation}
\begin{array}{lll}
    |\psi(t)\rangle & = & \displaystyle{d_0|n\rangle_x+\sum_{n=1}^\infty\sum_{s=\pm}d_{n\pm}e^{\pm i\Omega_n(t)}|n\rangle_{x\pm}}\\ 
    & = & \displaystyle{d_0|n\rangle_x+\sum_{n=1}^\infty\sum_{s=g,e}c_{ns}(t)|n,s\rangle,}
    \end{array}
\end{equation}
with the time-dependent ``Rabi frequency'' $\Omega_n(t)=\sqrt{n}g_0\sin(\omega t)/\omega$, and 
\begin{equation}
\begin{array}{l}
c_{ne}(t)=\frac{1}{\sqrt{2}}\left(d_{n+}e^{i\sqrt{n}\sin(\omega t)/\omega}+d_{n-}e^{- i\sqrt{n}\sin(\omega t)/\omega}\right),\\
c_{ng}(t)=\frac{1}{\sqrt{2}}\left(d_{n+}e^{i\sqrt{n}\sin(\omega t)/\omega}-d_{n-}e^{- i\sqrt{n}\sin(\omega t)/\omega}\right).
\end{array}
\end{equation}
If, for example, the atom is initially in its excited state, $c_{ng}(0)=0$, we have $d_{n+}=d_{n-}\equiv d_n$. If the photon field is initially in a coherent state with an amplitude $\alpha$, we further have $\sqrt{2}d_n=e^{-|\alpha|^2}\frac{\alpha^n}{\sqrt{n!}}$. The atomic inversion reads~\cite{prants1992jaynes,joshi1993generalized} 
\begin{equation}\label{atinv}
    W(t)\equiv\langle\hat\sigma_z\rangle=\sum_n2|d_n|^2\cos\left[2\Omega_{n+1}(t)\right].
\end{equation} 
As for the regular JC model~\cite{meystre1973destruction,narozhny1981coherence}, for very slow driving ($\omega\ll1$) the above inversion will display the familiar pattern of collapses and revivals as shown in fig.~\ref{fig1}. What is different from the regular JC model is the sinusoidal dependence of the Rabi frequency $\Omega_n(t)$. Initially the argument of the cosine function increases from zero, exactly like the time-independent JC model (where the argument of the cosine $g\sqrt{n}t$), but contrary to that model where the argument grows indefinitely, in our model it starts to decrease after $t=t_\mathrm{tr}=\pi/2\omega$. After a time  $2t_\mathrm{tr}$ the sinusoidal is again zero and we must have that every term in the sum~(\ref{atinv}) return back in phase leading to a perfect revival. The sinusoidal time dependence effectively reverse the evolution and clearly, this behaviour repeats as time progresses. The first revival time of the JC model is $t_\mathrm{r}=2\pi|\alpha|/g$~\cite{larson2022jaynes}, and hence, if $t_\mathrm{tr}<t_\mathrm{r}$ there is no time for the JC evolution to establish a revival, and the only revival will be the one stemming from the effective time reversal at $2t_\mathrm{tr}$. On the other hand, if $t_\mathrm{tr}>t_\mathrm{r}$ the system will display at least one revival before time is effective reversed. We should point out that the effective time-reversal evolution appears from considering the resonant case $\Delta=0$, while for $\Delta\neq0$ this effect is lost.

There are two additional time scales in the problem; the inverse Rabi frequency $t_\mathrm{R}=\pi/g|\alpha|$ and the collapse time $t_\mathrm{c}=1/2g$~\cite{larson2022jaynes}, and in principle $t_\mathrm{tr}$ should also be compared to these. For example, for very rapid driving, $\omega\gg g$ the inversion will be hindered. To lowest order such dynamical freezing can be understood from time-averaging the Hamiltonian which would result in a decoupling of the atom from the field. However, more insight is given by performing a second rotating wave approximation of the Hamiltonian~(\ref{tjc}) to find
\begin{equation}\label{tjc2}
    \hat h_n\approx\left[
    \begin{array}{cc}
    \frac{\Delta-\omega}{2} & \frac{g_0}{2}\sqrt{n}\\
    \frac{g_0}{2}\sqrt{n} & -\frac{\Delta-\omega}{2}
    \end{array}
    \right].
\end{equation}

Let us now take the alternative approach and analyze the problem in terms of Floquet theory by utilizing the mapping~(\ref{fmap}). For simplicity we consider the Fock states $|n\rangle_{x\pm}$, such that the Hamiltonian is diagonal within the two-level structure, and the Floquet Hamiltonian becomes
\begin{equation}\label{wsham}
    \hat H_\mathrm{JC}^\mathrm{F}(n)=\omega\hat E_0+\frac{g\sqrt{n}}{2}\left(\hat E^++\hat E^-\right)\hat\tau_z,
\end{equation}
where $\hat\tau_z|n\rangle_{x\pm}=\pm|n\rangle_{x\pm}$. We recall that $n$ in the above equation represents the photon number. In the appendix~\ref{appA} we discuss the idea of FSLs~\cite{saugmann2023fock}. As done in the appendix, we can generalize FSLs to a {\it Floquet FSL}; the composition of the Floquet space $\mathcal{F}=\mathcal{H\otimes T}$ results in an extended FSL where harmonic drive provides an additional dimension to the lattice. In the present model, we see that each copy of the Hamiltonian~(\ref{wsham}) is a Wannier-Stark Hamiltonian~\cite{hartmann2004dynamics}. Thus, the Floquet FSL consists of an infinite set of (decoupled) 1D chains -- each chain represents a specific photon number $n$. We may note that by adding a photon drive $\eta\left(\hat a+\hat a^\dagger\right)$, the photon number would no longer be preserved and the chains would be coupled, resulting in a two square lattices. 

The energies $\varepsilon_j$ and eigenstates $|\phi_j(n)\rangle$ of the Hamiltonian can be found analytically~\cite{fukuyama1973tightly}
\begin{equation}
    \varepsilon_j=j\omega,\hspace{1cm}|\phi_j(n)\rangle=\sum_m J_{l-m}\left(\pm\frac{g\sqrt{n}}{\omega}\right)|m\rangle.
\end{equation}
Here, $J_{m-j}(\pm g\sqrt{n}/\omega)$ is the Bessel function, which is predominantly localized in the interval $|m-j|<g\sqrt{n}/\omega$, and the $\pm$-sign stems from the value of $\hat\tau_z$ (for the Bessel functions $J_j(x)$ one has that they are even/odd for even/odd $j$). Note how the {\it Wannier-Stark ladder spectrum} $\varepsilon_j$, i.e. the Floquet quasi energies, is equidistant and does {\bf not} depend on the coupling $g\sqrt{n}$. This is a result from the lattice size extending to infinity, while in numerics, when a trucnation is imposed, one finds a slight dependence. 

\begin{figure}[ht!]
\centering\includegraphics[width=8.3cm]{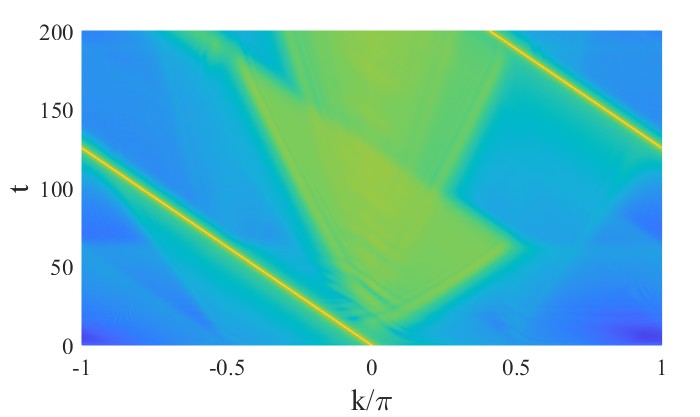}
\caption{The logarithm of the Floquet `momentum' distribution, $\log\left[P(k,t)\right]$, corresponding to the simulation of fig.~\ref{fig1}. Initially, the Floquet state is the phase state $|\theta=0\rangle$ or equivalently the $k=0$ quasi-momentum state, and with time it evolves as $k=-\omega t$ in accordance to Bloch oscillation. The population of the center light regions are actually greatly exaggerated by plotting the logarithm; this is actually a vanishingly small part of the total population. The Floqquet lattice for this numerics comprises 1401 sites, and by increasing this the center-most population is suppressed.  }\label{fig1a}
\end{figure}

Given the spectrum and eigenstates of the Floquet Hamiltonian, it is possible to derive analytically the time-dependence of various expectations~\cite{hartmann2004dynamics}. It is known that the time-evolution generated by the Wannier-Stark Hamiltonian $\hat H_\mathrm{JC}^\mathrm{F}$ results in periodic {\it Bloch oscillations} if the state is initially localized in momentum space, while if is is initially localized in real space you find instead periodic {\it breathing modes} where the state repeatedly broadens and contracts. However, we must recall that our initial state in the $\mathcal{T}$ subspace should be a phase state $|\theta\rangle$, which is extended over the whole Floquet lattice. Thereby, such oscillations are not clearly manifested in real space (see, however, next section where they show up in the numerics when a truncation has been considered). One the other hand, in reciprocal space a phase state has a definite `momentum' $k=\theta$. For the Wannier-Stark Hamiltonian, the quasi-momentum obeys $k(t)=k_0-\omega t$~\cite{hartmann2004dynamics}, and hence we expect a linear growth as time evolves. Of course, the quasi-momentum is limited to the first Brillouin zone $-\pi\leq k<\pi$, and over longer times it repeatedly sweeps the Brillouin zone. We demonstrate this is fig.~\ref{fig1a} by plotting the `momentum' distribution $P(k,t)$ which is obtained by Fourier transforming the time-evolved state of the $\mathcal{T}$ space, i.e. we first need to integrate out the photon and spin/atom degrees of freedom. When simulating the systems numerically we have to truncate the infinite state space. The initial state of the photon mode will be a coherent state, $|\alpha=5\rangle$, and for the atom it will be $|g\rangle$. For the simulation of the Floquet Hamiltonian we will take the initial state as the ground state of the ``bare'' Floquet Hamiltonian $\hat H_0^\mathrm{F}=\frac{\Delta}{2}\hat\sigma_z-\left(\hat E^++\hat E^-\right)$, where we impose periodic boundary conditions of the Floquet space, i.e. the phase state $|\theta=0\rangle$. Hence, in the reciprocal space this is the $k=0$ quasi-momentum state. The figure makes clear how the system shows the characteristic Bloch oscillation behaviour. The effect is somewhat less clear since we plot the logarithm of the distribution. We do this in order to envision the effects from the Floquet space truncation, which here show up as populated regions around the $k=0$ quasi-momentum. Without the logarithm these are hardly visible by bare eye.

 While fig.~\ref{fig1a} expose the characteristic time evolution of Wannier-Stark problem, the signatures of the space truncation is only clear from the fact that the logarithm, rather than the true, of the distribution is presented. To explore this numerical approximation we show the result of the atomic inversion, again with a finite Floquet lattice size of size 1401, in fig.~\ref{fig1} to be compared with the exact result. At short times the agreement is good, and the initial collapse and first revival is evident. However, for longer times the shortings of the state space truncation sets in and the finite size Floquet analysis qualitatively fails. This time can be pushed by increasing the lattice size, but it rapidly becomes computationally costly.

\section{Driven Lipkin-Meshkov-Glick model}\label{lmgsec}
The Lipkin-Meshkov-Glick model (see also Appendix A), first introduced in the nuclear physics community as a solvable many-body model~\cite{lipkin1965validity}, contains a single large spin $S$. The physics involves an interplay between a term attempting to align the spin in the $z$ direction and a coupling of the $x$ spin component. Specifically, for a negative ferromagnetic coupling, it becomes critical in the thermodynamic limit of $S\rightarrow\infty$~\cite{latorre2005entanglement,dusuel2004finite,ribeiro2007thermodynamical,larson2010circuit}. It can be viewed as a transverse field Ising model with an ``all-to-all'' coupling, where each spin couples identically to every other spin. Furthermore, since the total spin is conserved, every spin symmetry sector realizes a spin-$S$ LMG model. As expected from such a long-range model, the critical behavior belongs to the mean-field universality class [30].

This model has garnered significant attention in the quantum optics community~\cite{larson2017some}, partly because it results from eliminating the boson degrees of freedom of either the well-known Tavis-Cummings or the Dicke models~\cite{morrison2008dynamical,larson2010circuit}. It has also been considered when describing the bosonic Josephson effect~\cite{milburn1997quantum}.

The periodically driven LMG model, albeit in a different setting, has undergone thorough analysis dating back to the 1980s~\cite{frahm1985dynamics}. More precisely, the so-called ``kicked top'' is a specific realization of the driven LMG model. In the case of periodic kicking, either of the two terms in the Hamiltonian becomes time-dependent as $\delta(t-n\tau)$ (for an integer $n$, and some period $\tau$). The delta time dependence allows for analytical explorations, and the kicked top is particularly interesting as it demonstrates a system transitioning from regular to chaotic behavior as the kicking strength is increased~\cite{ghose2008chaos,chaudhury2009quantum,mumford2023many}. It also serves as a platform for studying dynamical localization, where ergodicity can be lost due to a dynamical analogue of Anderson localization~\cite{haake1991quantum,wang2023statistics}. These observations have led to the consideration of the model as a candidate for implementing time crystals~\cite{russomanno2017floquet}.

The harmonically driven LMG model, as considered in this paper, is not as easily analyzed analytically without invoking any approximations. In the past, it has been examined in terms of dynamical criticality~\cite{engelhardt2013ac}. We note that criticality can manifest in various ways in non-equilibrium situations, as briefly discussed in~\cite{bento2023krylov} in terms of the LMG model. 

While there are various versions of the LMG model, we will focus on the driven LMG model given by the following Hamiltonian:
\begin{equation}\label{lmgham}
    \hat H_\mathrm{LMG}=\Delta\cos(\omega t)\hat S_z-\frac{\lambda}{S}\hat S_x^2.
\end{equation}
Here, $\Delta$ represents the effective drive amplitude, $\omega$ is the drive frequency as before, $\lambda$ is the coupling strength, and $S$ is the total spin. The spin operators obey the regular commutators $\left[\hat S_i,\hat S_j\right]=i\varepsilon_{ijk}\hat S_k$, with $\varepsilon_{ijk}$ being the Levi-Chivita antisymmetric tensor.

In contrast to the driven JC model discussed in the previous section, here the drive is applied to the ``field'' and not to the coupling. The simple reason for this choice is that it leads to more interesting dynamical structures. In ref.~\cite{engelhardt2013ac} (similarly in~\cite{bastidas2012nonequilibrium,dasgupta2015phase,das2023periodically} for the Dicke model), the periodically driven LMG model was considered in a regime where a RWA was assumed. The RWA neglects rapidly rotating terms, analogous to a time averaging over the short time scale, and the effective Hamiltonian becomes time-independent. Within this regime, different phases were identified by introducing a mean-field energy and analyzing its local minima (which correspond to meta-stable states).

\begin{figure}[ht!]
\centering\includegraphics[width=8.3cm]{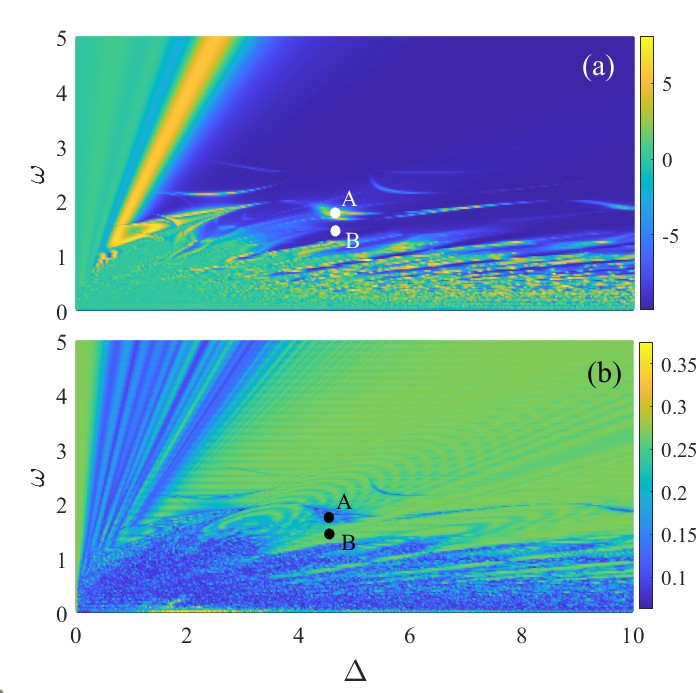}
\caption{Time-averaged magnitization $\langle\hat S_z\rangle$ (a) and phase space localization $PR$ (b) of the periodically driven LMG model~(\ref{lmgham}) as a function of the field strength $\Delta$ and drive frequency $\omega$. Several features stand out, such as high localization in the state for rapid and strong driving, with exceptions highlighted for example by the point A. A slow driving frequency tends to thermalize the system, similarly for a small $\Delta.$. The two dots mark the parameter values used in the following figure~\ref{lmgfig3}. The total spin $S=10$, and the system is initialized in the $|S,-S\rangle$ state.}\label{lmgfig1}
\end{figure}

We will not impose any RWA, and thereby not derive an effective time-independent Hamiltonian, but analyze the full problem numerically. The quantities of interest will typically be time-averaged, for example the magnetization takes the form
\begin{equation}
    S_z=\lim_{T\rightarrow\infty}\frac{1}{T}\int_0^T\langle\psi(t)|\hat S_z|\psi(t)\rangle dt.
\end{equation}
Another quantity of interest will be the phase space ``partition ratio''
\begin{equation}
    P\!R=\int Q^2(\theta,\phi)\,d\theta d\phi.
\end{equation}
Here, $Q(\theta,\phi)=\langle\theta,\phi|\hat\rho|\theta,\phi\rangle$ is the Husimi $Q$ function, and $|\theta,\phi\rangle=\exp\left[i\theta\left(\hat S_x\cos\phi-\hat S_y\sin\phi\right)\right]|S,S\rangle$ is a spin-coherent state with $|S,S\rangle$ a Dicke state~\cite{zhang1990coherent}. Often the inverse partition, $1/P$ ratio is instead considered, and more common is to define the partition ratio from a state $|\psi\rangle$\footnote{For some basis $\{|\varphi_n\rangle\}$, the partition ratio would be $P=\sum_n|\langle\varphi_n|\psi\rangle|^4$.} rather than the phase space distribution. $P$, as defined here, serves as a measure of phase space localization~\cite{evers2008anderson}. The closer its value is to unity, the more localized the state, while a smaller value indicates greater delocalization in phase space. For completeness, it is worth noting that the partition ratio is expressed in terms of the square of the Husimi distribution. However, other powers could also be considered, as is done when analyzing multi-fractal properties. Additionally, $P$ is related to the R\'enyi-Wehrl entropy, which has undergone thorough analysis in terms of the Dicke model in~\cite{romera2012husimi}.

\begin{figure}[ht!]
\centering\includegraphics[width=8.3cm]{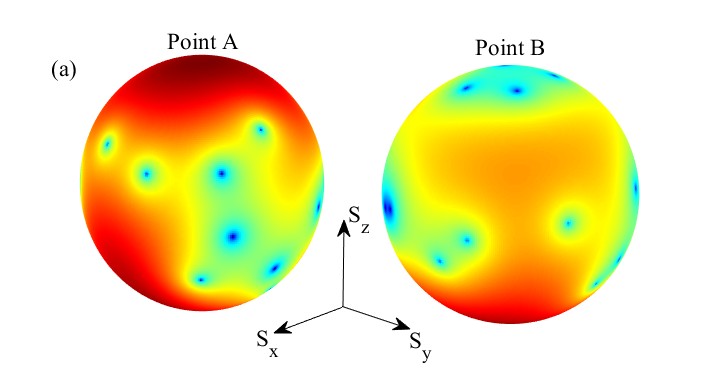}
\centering\includegraphics[width=8.3cm]{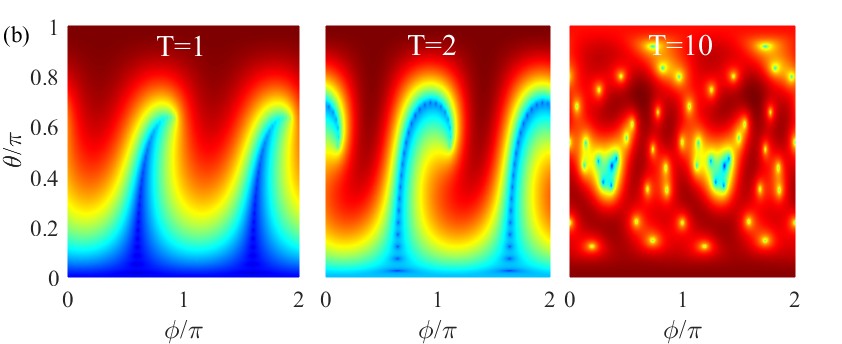}
\caption{Fig. 4. Upper plot (a): two examples of the Husimi distributions (more precisely $\log\left[Q(\theta,\phi)\right]$ to demonstrate localization) at a given time $T=200$, corresponding to points A and B in fig.~\ref{lmgfig1}. The final time $T$ influences the Husimi distributions, but the level of localization remains roughly constant. In the left plot (point A), the partition ratio is low, and the distribution is more delocalized (sharp red) than in the right plot (point B). The total spin is as before $S=10$. Lower plot (b): three snapshots ($T=1,\,2,\,10$) of $\log\left[Q(\theta,\phi)\right]$ in the $\theta\phi$-plane, for parameters $\Delta=\omega=1/2$ and $S=25$. Initially, the Husimi distribution has no zeros (the $|S,-S\rangle$ state), but as the state evolves, a string of $Z=2S$ zeros develops. For short times, the zeros (blue or yellow dots) follow nice curves, but already at $t=10$, the zeros are more or less randomly distributed, indicating chaotic behavior. }\label{lmgfig2}
\end{figure}

In fig.~\ref{lmgfig1}, we present the numerical results for the time-averaged magnetization $S_z$ (a) and the time-averaged partition ratio $P\!R$ (b) as functions of both the drive frequency $\omega$ and field amplitude $\Delta$. We numerically integrated the Schr\"odinger equation, as generated by $\hat H_\mathrm{LMG}$, over extended times (sufficient for convergence) and averaged the results. The plots illustrate a direct correlation between magnetization and phase space localization. For small $\omega$, the system is ergodic, and its state populates large fractions of phase space. Large delocalization is also observed for large $\omega$'s and moderate $\Delta$'s, but this does not appear to be a chaotic regime, as indicated by the regular pattern of $P\!R$ for these parameters. For large $\omega$ and $\Delta$, the system is highly magnetized and localized to its phase space south pole.    

It is observed that small parameter fluctuations can significantly alter the behavior of the evolution, as exemplified by the two points A and B in fig.~\ref{lmgfig1}. To take a closer look at these points, we plot the logarithm of the Husimi distributions $Q(\theta,\phi)$ in fig.~\ref{lmgfig2} (a) for these two sets of parameters. The sharper red regions are more extended for case A, indicating a more delocalized state as anticipated. As time progresses, and given a sufficiently long duration, the distributions change, while the number of zeros $Z$ remains constant, and the $P\!R$ stays more or less constant as well.

A striking feature of the shown Husimi distributions is its zeros. 
Analysis of the zeros of the Husimi distribution has a rather long history~\cite{leboeuf1990chaos,cibils1992zeros,korsch1997zeros,romera2012husimi}. It is related to the {\it stellar representation} of quantum states, i.e. the zeros uniquely determines the state. For spin states it is related to the {\it Majorana representation}~\cite{serrano2020majorana}. It is known that distribution of these zeros and the number $Z$ of them can be considered as an indicator of chaotic evolution. For regular motion the zeros tend to align in regular structures, while if the system displays chaos the zeros appear randomly spread over phase space. For spin $S$ states, the Majorana representation of a state is a $2S$ polynomial, from which it follows that the Husimi distribution can at most have $Z=2S$ zeros~\cite{amiet1999coherent,serrano2020majorana,chabaud2020stellar}. 

In fig.~\ref{lmgfig2} (b), we illustrate how these zeros evolve for one particular Husimi distribution. Initially, there are no zeros, and as the state evolves, two lines of zeros develops from the south pole. These curves of zeros twist, and rapidly the zeros depart from the line, seemingly randomly spreading over the full phase space. This clearly demonstrates chaotic behavior. We may mention (although not shown) that if we let $\omega=0$, we regain the time-independent LMG model, which is known to be integrable. In that case, one also observes the buildup of $Z=2S$ zeros, but this time they appear in regular patterns.

\begin{figure}[ht!]
\centering\includegraphics[width=8.3cm]{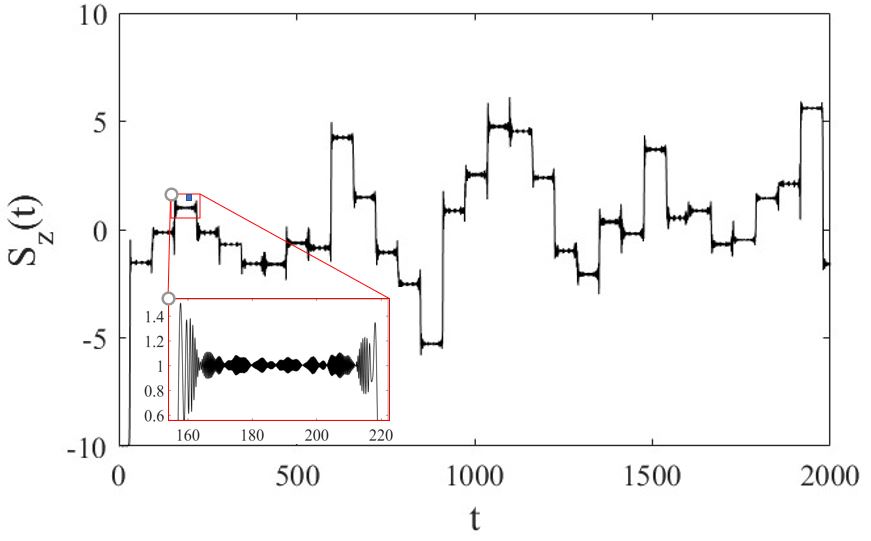}
\caption{Time evolution of the magnetization $S_z(t)=\langle\hat S_z\rangle$ for the same initial state as in the previous figures, and this time for the dimensionless parameters $\Delta=20$, $\omega=0.05$, and $S=10$. At periodic times, determined by $\cos(\omega t)=0$, the magnetization suddenly make a jump and get ``stuck'' at a random new magnetization where it rapidly oscillates around, as explicitly shown in the inset.   }\label{lmgfig3}
\end{figure}

In sec.~\ref{jcsec}, we discussed how the mixing of different time scales alters the evolution. Naturally, the same applies here. For very slow driving, $\omega\ll1$, we may encounter adiabatic time evolution as long as $S$ is not too large. The criticality of the LMG model implies that adiabaticity is doomed in the thermodynamic limit since the system is gapless at the critical points. For rapid driving, the evolution is approximately diabatic. This is the regime in the upper right corner of fig.~\ref{lmgfig1}. When $|\Delta|\gg\omega$, the intrinsic LMG time scale is short in comparison to that of the drive. Here, the chaotic evolution manifests in a generic feature, as we demonstrate in fig.~\ref{lmgfig3} by plotting the time evolution of the magnetization $S_z$ (hence, not time-averaged). For $\cos(\omega t)$ away from zero, as the inset shows, the magnetization oscillates with a frequency set by the instantaneous ``field strength'' $\Delta\cos(\omega t)S$. Whenever $\cos(\omega t)\rightarrow0$, the magnetization makes a jump to some random new magnetization, where it starts to oscillate around. For a very large spin $S$, a simple numerical exploration suggests that these jumps are indeed random~\footnote{We integrated the semi-classical equations of motion over long periods and collected the magnetizations where the system get stuck. Up to some boundary deviations, the distribution of these was found to be evenly distributed within the interval $[-S,S]$.}. Thus, the system can serve as a chaos-driven random number generator~\cite{yu2019survey}. The jumps originate from multi-level {\it Landau-Zener transitions}~\cite{landau1932theorie,zener1932non,larson2014interaction}. It is known that Landau-Zener transitions are highly sensitive to dynamical phases, and the more involved levels this will generate seemingly random transitions. It is interesting to note that the same random structure was found in an optomechanical system in~\cite{larson2011photonic}. In that survey, the two time scales were both intrinsic, while in the present model, one is set by the classical drive. 

\begin{figure}[ht!]
\centering\includegraphics[width=8cm]{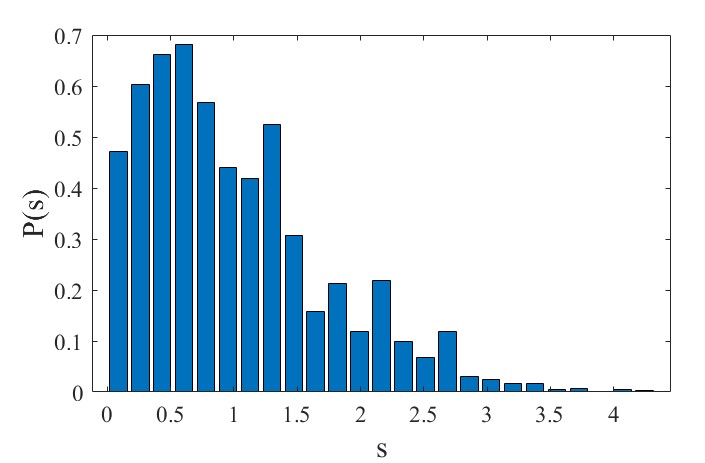}
\caption{The level statistics of the Floquet Hamiltonian~(\ref{flmgham}), for a Floquet lattice of 201 sites, $\Delta=2$, $\omega=0.25$, and $S=10$. Although the distribution is neither Poissonian nor of Wigner-Dyson shape, it clearly exhibits level repulsion.} \label{lmgfig5}
\end{figure}

Up till now, all numerical simulations have been performed for the time-dependent Hamiltonian~(\ref{lmgham}). We have not analyzed the problem from the perspective of its Floquet Hamiltonian, which in this case becomes  
\begin{equation}\label{flmgham}
    \hat H_\mathrm{LMG}^\mathrm{F}=\omega\hat E_0+\frac{\Delta}{2}\left(\hat E^++\hat E^-\right)\hat S_z-\frac{\lambda}{S}\hat S_x^2.
\end{equation}
In terms of a Floquet FSL, see the appendix, this Hamiltonian represents a tilted square lattice. One dimension corresponds to the Floquet space, and the other, which is finite, to the spin space. The clear advantage of the Floquet Hamiltonian is that it is time-independent. However, it increases the state space, leading to truncation approximation in numerical computations.

It actually turns out to be far more computationally costly to solve the Floquet problem than the time-dependent one. The main reason for this is that one must consider very large systems in order to correctly capture the exact dynamics, which is understood from the fact that our initial state is a phase state that is delocalized in the Floquet FSL. Nevertheless, insight into the physics can be gained by analyzing the Floquet Hamiltonian.

One question that comes to mind is how the chaos is manifested in $\hat H_\mathrm{LMG}^\mathrm{F}$. The Hamiltonian bears similarities with the Dicke model, which is known to display a transition from regular to chaotic evolution across the normal-superradiant phase transition~\cite{emary2003chaos}. What differs between $\hat H_\mathrm{LMG}^\mathrm{F}$ and the Dicke Hamiltonian is that here a spin couples to the `Euclidean' field (represented by the $\hat E$-operators) rather than a boson field, and the bare spin Hamiltonian is quadratic ($\hat S_X^2$ instead of $\hat S_x$). Chaos in the Dicke model has been analyzed in various ways, for example, by the {\it level statistics}~\cite{haake1991quantum}.

Given the spectrum $E_n$ (ordered $E_1\leq E_2\leq E_2\leq\dots$) we introduce the level spacings $S_n=E_{n+1}-E_n$. With $S_n$, one constructs the normalized distribution function $P(S)$\footnote{The level spacing $S$ should not be confused with the spin $S$.}. In case of an integrable Hamiltonian we expect a {\it Poisson distribution}
\begin{equation}
    P_\mathrm{P}(S)=e^{-S},
\end{equation}
while for chaotic motion we expect a {\it Wigner surmise}
\begin{equation}
    P_\mathrm{W}(S)=a_\beta S^\beta e^{-b_\beta S^2},
\end{equation}
where $a_\beta$ and $b_\beta$ are constants and the exponent $\beta$ is set by the symmetry classes~\cite{haake1991quantum}. The fact that $\lim_{S\rightarrow0}P_\mathrm{W}(S)=0$ implies that that the energy levels do not cross, something referred to as {\it level repulsion}. 

Taking the Wigner surmise as a chaos indicator, one should note that fully developed chaos is not to be expected for low energies. Therefore, one typically considers intermediate energies. For a finite-dimensional system, the central part of the spectrum is used. Furthermore, the spectrum is typically {\it unfolded} before calculating the distribution function. This involves normalizing the level distances $S$ relative to a mean distance~\cite{haake1991quantum}. The procedure of unfolding follows these steps:
\begin{enumerate}
    \item Compute $N(E)$, which counts the number of states with energies smaller or equal to $E$.
    \item Fit a smooth curve $R(E)$ to the step-like function $N(E)$ (we use a seventh order polynomial).
    \item Identifies energies $e_n=R(E_n)$, and calculate the corresponding `unfolded' distribution function $P(s)$. 
\end{enumerate}

An example of the distribution function P(s)P(s) is shown in fig.~\ref{lmgfig5}. It demonstrates level repulsion, as one would expect for chaotic evolution, but it is not fully developed. Part of this could be explained by the fact that we consider the full spectrum without removing degeneracies arising from symmetries. The Hamiltonian $\hat H_\mathrm{LMG}^\mathrm{F}$ possess two $\mathcal{Z}_2$-parity symmetries, $(\hat E_0,\hat E^\pm,\hat S_x,\hat S_y,\hat S_z)\rightarrow(\hat E_0,-\hat E^\pm,-\hat S_x,\hat S_y,-\hat S_z)$ and $(\hat E_0,\hat E^\pm,\hat S_x,\hat S_y,\hat S_z)\rightarrow(\hat E_0,-\hat E^\pm,\hat S_x,-\hat S_y,-\hat S_z)$. We should also mention that the actual shape of $P(s)$ depends strongly on the system parameters, and sometimes the distribution looks more Poisson and sometimes nothing like Poisson or Wigner.

\begin{figure}[ht!]
\centering\includegraphics[width=8.3cm]{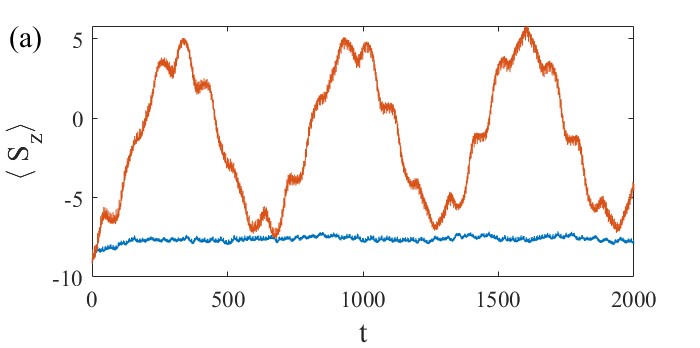}
\centering\includegraphics[width=8.3cm]{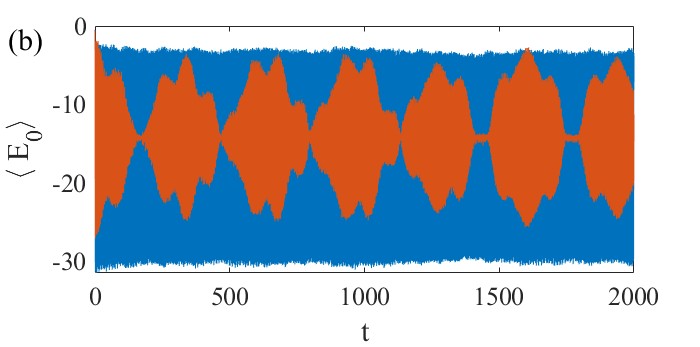}
\caption{Time evolution of the magnitization $\langle\hat S_z\rangle$ (a) and $\langle\hat E_0\rangle$ (b) for parameters corresponding to the case A (red curve) and B (blue curve) of fig.~\ref{lmgfig1}. In case B, the system remains highly magnetized, while in case A, the magnetization oscillates over large periods. In the lower plot, it is evident that in the Floquet space, the two cases behave qualitatively differently. Both curves show rapid oscillations corresponding to Bloch oscillations, but the red curve exhibits collapses and revivals on top of these. The number of lattice sites (states in the Floquet space) is taken to be 101. }\label{lmgfig6}
\end{figure}

Let us conclude our explorations by revisiting points A and B in fig.~\ref{lmgfig1}. Our initial Floquet state is delocalized over the full lattice, i.e., $\langle\hat E_0\rangle=0$. However, our truncation implies that  $\langle\hat E_0\rangle$ may actually display non-trivial evolution, as demonstrated in Fig. 7. In (a), we display the evolution of the magnetization $\langle\hat S_z\rangle$, where a clear difference is observed. For case B, the system rapidly settles to a certain magnetization and thereafter only shows small oscillations around it. It resembles rapid thermalization, but we should remember that the state is actually fairly localized in phase space here, as seen in fig.~\ref{lmgfig2}. One could speculate whether this could be a result of {\it dynamical localization}~\cite{haake1991quantum}, but this seems highly unlikely since the kicked top does not exhibit dynamical localization, and this should hold for a harmonically driven top as well. For case A, we observe large-amplitude sustained oscillations spanning over long periods. We believe that these are instances where constructive interference in the repeated multi-level Landau-Zener transitions occurs. An indicator supporting this is that by increasing $S$, it seems as if these 'resonances' become rarer.

Part of the evolution in the Floquet space is captured in fig.~\ref{lmgfig1} (b) by showing the evolution of the expectation $\langle\hat E_o\rangle$ for the two cases. Apart from short oscillations resulting from the tilt, i.e. manifestations of Bloch oscillations, case A also displays a collapse-revival pattern. For one oscillation of $\langle\hat S_z\rangle$, $\langle\hat E_0\rangle$ makes two oscillations. We should recall that the Floquet Hamiltonian is truncated, and at these long times we cannot expect convergence of the results. But even so, it is clear that the two cases behave very different.

\section{Conclusion}\label{con}
We have examined two periodically driven models, the JC and LMG models, which frequently appear in the quantum optics community. In both cases, the dynamics strongly depend on the involved time scales: those set by the drive and the intrinsic ones. In the JC model, multiple intrinsic time scales exist, such as Rabi oscillations, collapse, and revival times. When the time scale of the drive is comparable to the Rabi time, a rather complex dynamics emerges. Conversely, if this time scale is comparable to the revival time, the dynamics can be understood from an adiabatic perspective. In such cases, the drive can act as an effective reversal of time, akin to spin-echoes. 

In our Floquet analysis of the problem, we introduce a Floquet FSL, which are generalizations of FSLs for time-independent models~\cite{wang2016mesoscopic,cai2021topological,saugmann2023fock,yuan2024quantum}. The time-independent Floquet Hamiltonian determines the Floquet lattice in the same way as for regular FSLs, with the Floquet space adding an additional dimension to the lattice. For the JC model, counterparts of Bloch oscillations are found in the Floquet FSL.

The LMG model becomes chaotic when a periodic drive is included, giving rise to several novel features. While the undriven LMG model is not chaotic, it is not immediately clear whether the driven one should generally lead to heating and ergodicity. Indeed, the kicked LMG has been studied in terms of Floquet time-crystals, which evade heating up~\cite{russomanno2017floquet}. Despite the states being fairly localized in phase space, we do observe several signatures of chaos, such as the distribution of zeros of the Husimi function and random spin jumps. Many of these results can be analyzed in terms of repeated multi-level Landau-Zener transitions, where strong chaos leading to ergodicity arises from complex multi-level interferences. The random jumps occur for slow driving, but not during fully adiabatic evolution. Additionally, the interferences may lead to `resonances' where the system behaves qualitatively differently for slight parameter changes. These resonances also manifest in the Floquet FSL in the form of collapses and revivals.

We have limited our analysis to two models, but one could envision exploring many other models as well. It would be intriguing to investigate different Floquet FSL geometries. In principle, by employing two harmonic drives with incommensurate frequencies, it is possible to create Floquet FSLs with two additional dimensions, one for each drive. However, a drawback of the Floquet FSL, compared to regular FSLs, is that the initial states are delocalized. This aspect makes them less practical for "lattice engineering" and simulations involving synthetic dimensions. 

The LMG model is integrable, unlike the Dicke model, for instance. Consequently, one might anticipate qualitatively different behaviors for a driven Dicke model. For example, it could exhibit dynamical localization, which differs from the localization discussed in~\cite{wang2020statistical}.

\appendix
\section{Fock state lattices}\label{appA}
In this appendix we briefly discuss Fock state lattices and consider it in terms of the LMG model. The FSL for generalized JC models was first introduced in~\cite{wang2016mesoscopic}. Each Fock state represents a quantum state of one particular lattice site. Tunneling between the synthetic lattice sites occurs via the action of the ladder operators, but due to the square-root normalization, e.g. for bosons $\hat a|n\rangle=\sqrt{n}|n\rangle$, the FSL will typically not be translationally invariant even within the bulk. Nevertheless, as demonstrated in~\cite{cai2021topological} for the three-mode JC model and in~\cite{saugmann2023fock} for various other setups, lattice properties like topology may survive. Most recently, the topological aspects has been experimentally explored~\cite{deng2022observing}.

Characteristic of many models of light-matter interaction, as found in the quantum optics setting, is that they are typically restricted to few degrees of freedom. The JC model is a prime example~\cite{jaynes1963comparison,shore1993jaynes,larson2022jaynes}. Furthermore, stemming from dipole interactions, the interactions are usually linear in the boson and spin degrees of freedom, meaning that in the Fock basis~(\ref{fockst}), the Hamiltonian becomes sparse with a structure similar to tight-binding lattice models. In~\cite{saugmann2023fock}, this was discussed for numerous models, and it was demonstrated that most known light-matter interaction models in the quantum optics community resulted in FSLs with the same geometry as well-known lattice models found in solid-state physics.

 For clarity, let us explore the time-independent example of sec.~\ref{lmgsec}, namely the LMG Hamiltonian~(\ref{lmgham}) that takes the form
 \begin{equation}\label{lmgham2}
     \hat H_\mathrm{LMG}=\Delta\hat S_z-\frac{\lambda}{S}\hat S_x^2.
 \end{equation}
The model supports a $\mathbb{Z}_2$-parity symmetry characterized by $\left(\hat S_x,\hat S_y,\hat S_z\right)\rightarrow\left(-\hat S_x,-\hat S_y,\hat S_z\right)$, and furthermore the total spin $S$ is preserved. The denominator in the last term warrants that the two terms scale as $\mathcal{O}(S)$ in the large $S$ limit. As mentioned in the main text, in the thermodynamic limit, $S\rightarrow\infty$, the model is critical at the coupling strength $\lambda_c=|\Delta|$; below the critical point the ground state is ferromagnetic with $S_x=\langle\hat S_x\rangle=0$ (and $S_z\neq0$), while above it is paramagnetic $S_x\neq0$~\cite{dusuel2004finite,ribeiro2007thermodynamical,larson2010circuit,latorre2005entanglement}. 

\begin{figure}[ht!]
\centering\includegraphics[width=8cm]{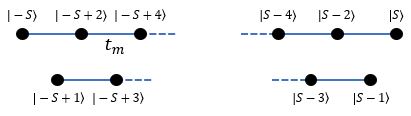}\label{lmgFSL}
\caption{The FSL of the LMG model~(\ref{lmgham2}). The parity symmetry divide the lattice into two 1D chains, one of even parity comprised of all states $|m\rangle$ with even quantum numbers $m$, and one with odd parity for the odd $m$'s. The blue lines connecting neighboring sites represent non-vanishing tunnelings. The corresponding tunneling rates $t_m$ are identified with the square-root prefactors given in eq.~(\ref{ladop}), i.e. $t_m=\alpha_{m+}\alpha_{m+1+}$.}
\end{figure}

The Fock states, according to~(\ref{fockst}), are $|m\rangle$ with $-S\leq m\leq S$, and 
\begin{equation}\label{ladop}
\begin{array}{l}
     \hat S_z|m\rangle=m|m\rangle, \\
     \hat S_x|m\rangle=\alpha_{m+}|m+1\rangle+\alpha_{m-}|m-1\rangle,
\end{array}
\end{equation}
with the normalization $\alpha_{m\pm}=\sqrt{S(S+1)-m(m\pm1)}$ (note that hermiticity implies $\alpha_{m+}=\alpha_{m+1-}$). The parity symmetry implies that the Hamiltonian couples states $|m\rangle$ with odd (even) $m$ to new states $|m'\rangle$ with odd (even) $m'$. Thus, we expect the FSL to be comprised of two 1D chains, as visualized in fig.~\ref{lmgFSL}. In the figure, we do not emphasize the onsite energies set by the diagonal terms of the Hamiltonian. In particular, the field $h\hat S_z$ induces a linear tilt, $hm$, of the lattice, while the interaction term causes an onsite shift $\alpha_{m+}^2+\alpha_{m-}^2=2S(S+1)+2m^2$. If the $\mathbb{Z}_2$-symmetry is broken, by adding for example a field $\lambda\hat S_x$ to the Hamiltonian~(\ref{lmgham2}), the two chains would be coupled and one instead finds a triangular ladder. 

For harmonically driven systems, the method outlined in this work allows the introduction of a Floquet FSL.  In particular, we identify the Fourier modes $m$ with lattice sites, i.e. $|m\rangle$ represents the state for a particle residing on site $m$. Hence, $\cos(\omega t)$ translates into a tight-binding Hamiltonian which induces hopping between neighbouring lattice sites. As a most simple example we can take the Rabi model with its corresponding Hamiltonian
\begin{equation}
    \hat H_\mathrm{R}(t)=\Omega\hat\sigma_z+g\cos(\omega t)\hat\sigma_x.
\end{equation}
Here, $\hat\sigma_\alpha$ ($\alpha=x,\,y,\,z$) are the Pauli matrices, $\Omega$ is the transition frequency of our two-level ``atom'', and $g$ is the amplitude of the periodic drive. The Floquet Hamiltonian takes the form~\cite{duan2023periodic}
\begin{equation}
    \hat H_\mathrm{R}^\mathrm{F}=\omega\hat E_0+\Omega\hat\sigma_z+\frac{g}{2}\left(\hat E^++\hat E^-\right)\hat\sigma_x.
\end{equation}
Following the notations of~(\ref{fockst}), the Fock states are $|s,m\rangle$. It is possible to give a physical picture of the various terms forming the Hamiltonian
\begin{itemize}
    \item The first term $\omega\hat E_0$ is diagonal in the Fock basis and thereby only shifts the onsite `energies'. This shift is linear and can be seen as a constant ``electric'' field for a charged particle, akin to the case of Bloch oscillations~\cite{dahan1996bloch,morandotti1999experimental,holthaus2000bloch} or the {\it Wannier-Stark ladder} problem~\cite{wilkinson1996observation,gluck1998calculation,hartmann2004dynamics}.
    \item The second term $\Omega\hat\sigma_z$ is again diagonal and provides an onsite shift depending on the internal spin state.
    \item The last term $\frac{g}{2}\left(\hat E^++\hat E^-\right)\hat\sigma_x$ represents the nearest neighbor tunneling between the $m$-sites. The tunneling is, however, accompanied by a spin-flip.
\end{itemize}

\bibliography{references}{}
\bibliographystyle{unsrtnat}

\end{document}